\documentclass[twocolumn,twoside,9pt]{IEEEtran}
\usepackage{bm}
\usepackage[final]{graphicx}
\newcommand{\bfa}[1]{{\bf #1}}

\newcommand{\E}[1]{E\left\{ #1 \right\}}
\newcommand{\Eu}[2]{{\rm E}_{#1}\left\{ #2 \right\}}

\newcommand{\N}[1]{{\mathcal N}\left(#1 \right)}

\renewcommand{\Pr}{{\textnormal P}}

\newcommand{\rhov}{{\bm \rho}}

\newcommand{\Av}{{\bf A}}

\newcommand{\bv}{{\bf b}}

\newcommand{\Hv}{{\bf H}}
\newcommand{\hv}{{\bf h}}
\newcommand{\Iv}{{\bf I}}

\newcommand{\nv}{{\bf n}}

\newcommand{\rv}{{\bf r}}
\newcommand{\Rv}{{\bf R}}
\newcommand{\sv}{{\bf s}}
\newcommand{\Sv}{{\bf S}}

\newcommand{\Vv}{{\bf V}}

\newcommand{\yv}{{\bf y}}

\newcommand{\lgth}{\stackrel{\textstyle>}{<}}
\newcommand{\defi}{\stackrel{\triangle}{=}}

\newtheorem{Lemma}{Lemma}
\newtheorem{PLemma}{Proof of Lemma}

\renewcommand{\(}{\left(}
\renewcommand{\)}{\right)}

\title{On The Tradeoff Between Two Types of Processing Gain}

\author{
Eran Fishler$^{\dag*}$ and H. Vincent Poor$^\ddag$
\thanks{This research was supported in part by the U.S. Army
Research Laboratory under Contract No. DAAD 19-01-2-0011, and in
part by the New Jersey Center for Wireless Telecommunications.}
\thanks{
   $^\dag$E. Fishler was with the Department of Electrical Engineering, Princeton University,
   Princeton, NJ. He is now with the Stern School of Buisness, New York University, NY, NY{\tt
   e-mail: ef485@stern.nyu.edu}
   }
\thanks{
   $^\ddag$H. V. Poor is with the Department of Electrical Engineering, Princeton University,
   Princeton, NJ 08544, USA, Tel: (609) 258-1816, Fax: (609) 258-1468,
   {\tt e-mail: poor@princeton.edu}
   }
\thanks{$*$ Corresponding author}
} 

\begin{document}

\maketitle

\begin{abstract}

One of the features characterizing almost every multiple access
(MA) communication system is the processing gain. Through the use
of spreading sequences, the processing gain of Random CDMA systems
(RCDMA), or any other CDMA systems, is devoted to both bandwidth
expansion and orthogonalization of the signals transmitted by
different users. Another type of multiple access system is Impulse
Radio (IR). IR systems promise to deliver high data rates over
ultra wideband (UWB) channels with low complexity transmitters and
receivers. In many aspects, IR systems are similar to time
division multiple access (TDMA) systems, and the processing gain
of IR systems represents the ratio between the actual transmission
time and the total time between two consecutive transmissions
(on-plus-off to on ratio). While CDMA systems, which constantly
excite the channel, rely on spreading sequences to orthogonalize
the signals transmitted by different users, IR systems transmit a
series of short pulses and the orthogonalization between the
signals transmitted by different users is achieved by the fact
that most of the pulses do not collide with each other at the
receiver.

In this paper, a general class of MA communication systems that
use both types of processing gain is presented, and both IR and
RCDMA systems are demonstrated to be two special cases of this
more general class of systems. The bit error rate (BER) of several
receivers as a function of the ratio between the two types of
processing gain is analyzed and compared under the constraint that
the total processing gain of the system is large and fixed. It is
demonstrated that in non inter-symbol interference (ISI) channels
there is no tradeoff between the two types of processing gain.
However, in ISI channels a tradeoff between the two types of
processing gain exists. In addition, the sub-optimality of RCDMA
systems in frequency selective channels is established.

\end{abstract}

%%%%%%%%%%%%%%%%%%%%%%%%%%%%%%%%%%%%%%%%%%%%%%%%%%%%%%%%%%
%%%%%%%%%%%%%%%%%%%%%%%%%%%%%%%%%%%%%%%%%%%%%%%%%%%%%%%%%%
%%%%%%%%%%%%%%%%%%%%%%%%%%%%%%%%%%%%%%%%%%%%%%%%%%%%%%%%%%
% Introduction
%%%%%%%%%%%%%%%%%%%%%%%%%%%%%%%%%%%%%%%%%%%%%%%%%%%%%%%%%%
%%%%%%%%%%%%%%%%%%%%%%%%%%%%%%%%%%%%%%%%%%%%%%%%%%%%%%%%%%
%%%%%%%%%%%%%%%%%%%%%%%%%%%%%%%%%%%%%%%%%%%%%%%%%%%%%%%%%%
\pagebreak

\section{Introduction}

%%%%%%%%%%%%%%%%%%%%%%%%%%%%%%%%%%%%%%%%%%%%%%%%%%%%%%%%%%
%%%%%%%%%%%%%%%%%%%%%%%%%%%%%%%%%%%%%%%%%%%%%%%%%%%%%%%%%%
% Motivation
%%%%%%%%%%%%%%%%%%%%%%%%%%%%%%%%%%%%%%%%%%%%%%%%%%%%%%%%%%
%%%%%%%%%%%%%%%%%%%%%%%%%%%%%%%%%%%%%%%%%%%%%%%%%%%%%%%%%%

\subsection{Motivation}

Multiple access (MA) communication systems are in widespread use.
It is enough to mention that almost every cellular phone system
and wireless local area network is an MA system. Many approaches
for implementing multiple access systems exist; for example direct
sequence code division multiple access (CDMA), frequency hopping,
and time division, to name a few \cite{Rappaport:01}. Recently,
Impulse Radio (IR) systems have been suggested as a simple way of
implementing MA systems \cite{Scholtz:93}. Impulse Radio systems
promise to deliver high data rates in multiple access channels
with low complexity transmitters and receivers. Currently, IR
systems are being considered for use in many applications and
mainly as the preferred solution for communication systems
transmitting over ultra wideband (UWB) channels
\cite{Cassioli:02,Cramer:02,Win:00,Win:02,Win:98a,Win:98b}.

The two most popular ways for implementing MA systems are CDMA and
time division multiple access (TDMA). These two types of systems
are based on two different, and even ``orthogonal" ideas. Consider
a CDMA system and a TDMA system assigned with equal bandwidth and
supporting identical users. In the CDMA system, each user's
transmitted signal's bandwidth is expanded using a distinct
spreading sequence, and all the users transmit simultaneously over
the same channel. The purposes of the spreading sequences are to
spread the transmitted energy over the assigned bandwidth and to
make different users' transmitted signals as close to orthogonal
as possible. Alternatively, in a TDMA system, each user transmits
for only a small fraction of the time but at a high data rate (and
hence the need for large bandwidth). By preventing simultaneous
transmissions from two different users, collisions between the
signals transmitted by different users are avoided and hence the
required rate from each user is achieved.

One type of CDMA systems are long-code CDMA systems, also known as
random CDMA (RCDMA) systems \cite{Tse:99,Verdu:99}. In RCDMA
systems each user uses a random spreading sequence for expanding
the bandwidth of the transmitted signal. Several MA communication
systems are based on long-code CDMA, with the IS-95 mobile phone
system being the most famous one \cite{Rappaport:01}. As mentioned
previously, IR systems have been suggested as a new approach to
implementing MA communication systems. IR systems transmit a
series of very short pulses, typically on the order of a fraction
of a nano-second in duration. Each user transmits each pulse at a
time slot randomly chosen, and each pulse is repeated several
times. The receiver, using appropriate signal processing
algorithms, recovers the transmitted bits \cite{Win:98}. IR
systems can be regarded as random TDMA systems where each user
transmits for a very short time at a time slot randomly chosen.

It is interesting to note that RCDMA systems and IR systems
represent two extremes of a wide range of MA systems. In the
first, the processing gain is devoted to increasing the signal
bandwidth and making the different users' transmitted signals as
close to orthogonal as possible, while in the second the
processing gain is mostly devoted to reducing the transmission
time, which in turn reduces the probability of several users
transmitting simultaneously. Although both IR and RCDMA systems
have been analyzed in the past, the literature still lacks a study
that examines the tradeoff between the types of processing gain
represented by these two systems. Moreover, systems that use both
types of processing gain have not been suggested and analyzed.
This paper offers such a study by examining the performance of IR
systems and the performance tradeoff between the two types of
processing gain as a function of the system parameters for a fixed
signaling environment.

The processing gain of IR systems, denoted hereafter by $N$, is
$N= N_f N_c$. The pulse rate, $N_f$, represents the first type of
processing gain, that is, the number of times each pulse is
repeated (either uncoded or coded). Alternatively, $N_c$, which is
the ratio between the average total time between two consecutive
transmissions and the actual transmission time, is the second type
of processing gain. Assume that the total processing gain is
fixed. By changing the pulse rate, $N_f$, the ratio between the
two types of processing gain is changed as well. The effect of
this change on the system's bit error rate (BER) is the main
interest of this paper. Thus, throughout the rest of the paper we
will compare systems that have equal {\em total} processing gain
but which divide this total processing gain between the two types
of processing gain differently.

%%%%%%%%%%%%%%%%%%%%%%%%%%%%%%%%%%%%%%%%%%%%%%%%%%%%%%%%%%
%%%%%%%%%%%%%%%%%%%%%%%%%%%%%%%%%%%%%%%%%%%%%%%%%%%%%%%%%%
% Signal model
%%%%%%%%%%%%%%%%%%%%%%%%%%%%%%%%%%%%%%%%%%%%%%%%%%%%%%%%%%
%%%%%%%%%%%%%%%%%%%%%%%%%%%%%%%%%%%%%%%%%%%%%%%%%%%%%%%%%%

\subsection{Signal Model}

Consider the case of downlink channel of a $K$-user TH-IR
synchronous cellular type system transmitting over a
frequency-flat channel. Note that generalizing the results
reported herein to more complex network configurations, e.g.,
uplink channels or asynchronous systems, can be easily carried out
at the expense of complicating most of the, already complicated,
computations, while the main results will essentially remain the
same. The received signal, of say the $k$th user (out of $K$ total
users in the system), in a binary phase-shift keyed random time
hopping impulse radio (TH-IR) system can be described by the
following continues time model,
   \begin{eqnarray}
      r(t) = \sum_{k=1}^K \sqrt{ \frac{E_k}{N_f} } \sum_{j=-\infty}^\infty
       d_j^k b_{\lfloor j/N_f
       \rfloor}^k w(t - j T_f - c_j^kT_c ) + n(t)
       \label{e. continues time received signal}
   \end{eqnarray}
where $T_f$ is the average pulse repetition time, $w(t)$ is the
transmitted unit-energy pulse, also referred to (in the UWB
literature) as the {\em monocycle}, and $E_k$ is the transmitted
energy per bit for user $k$. In order to allow the channel to be
exploited by many users and to avoid catastrophic collisions, a
long pseudo-random sequence $\{c_j^k\}$, such that $c_j^k$ is an
integer taking one of the values in $[0, 1, \ldots, N_c-1]$, is
assigned to each user. Each sequence, usually referred to as the
time hopping sequence, provides an additional time shift of
$c_j^kT_c$ seconds to the $j$th pulse of the $k$th user. In order
to avoid inter-pulse interference (IPI), it is usually required
that $T_c \leq \frac{T_f}{N_c}$, so that overlaps between pulses
originating from the same user are avoided. In typical IR systems
each data symbol is transmitted over a set of multiple monocycles
called a {\em frame}. Here $N_f$ denotes the number of pulses that
correspond to one information symbol, i.e., the number of
monocycles per frame. Thus, $b_{\lfloor j/N_f \rfloor}^k \in \{
\pm 1 \}$ is the information symbol transmitted during the $i =
\lfloor j/N_f \rfloor$th frame. $n(t)$ represent white, Gaussian
noise with power spectrum $\frac{N_0}{2}$ per hertz.

Two types of IR systems are considered in this paper. In the first
type $d_j^k = 1,\; \forall j,k$, while in the second the $d_j^k$'s
are binary random variables, independent for $(i,k)\neq(j,l)$,
taking each of the values $\pm 1$ with probability 1/2. The first
type of system was the first to be proposed in the literature for
transmission over UWB channels \cite{Martret:00a}, while a
different variant of the second was recently proposed in
\cite{Sadler:02}. In the sequel, the two types of systems are
referred to as {\em uncoded} and {\em coded} systems,
respectively. Note that a coded IR systems can model an RCDMA
system by taking $T_f = T_c$ and letting $N_f$ be the processing
gain of the RCDMA system.

Denote by $\{ s_j^k \}$ the following sequence
   \begin{eqnarray}
      s_j^k = \left\{ \begin{array}{cc}
         d_{\lfloor j/N_c \rfloor}^k & j -  N_f\lfloor j/N_c \rfloor = c_{\lfloor j/Nc\rfloor}^k \\
         0 & {\rm Otherwise}\\
         \end{array}\right..
         \label{e. spreading sequence model}
   \end{eqnarray}
The sequence $\{s_j^k\}$ can be regarded as a pulse (or chip) rate
spreading sequence where $s_j^k$ take the value $d_{\lfloor
j/N_c\rfloor}^k$ whenever a pulse is transmitted and zero
otherwise. Assuming, without loss of generality (wolog), that
$\frac{T_f}{T_c} = N_c$, the received signal can be described by
the following model,
   \begin{eqnarray}
       r(t)=\sum_{k=1}^K \sqrt{ \frac{E_k}{N_f} } \sum_{j=-\infty}^\infty s_j^k b_{\lfloor
       j/(N_fN_c) \rfloor}^k w(t - j T_c ) + n(t).
       \label{e. transmitted signal CDMA model}
   \end{eqnarray}

Although (\ref{e. transmitted signal CDMA model}) describes a
continuous time TH-IR signal, discrete time equivalent models for
CDMA systems can be used for describing the more general TH-IR
systems. Assume that the received signal is passed through a
linear filter matched to $w(t)$, and sampled at the pulse rate.
Denote by $\rv_i = [r_{i N_fN_c+1}\; \cdots \; r_{(i+1)
N_fN_c}]^T$ the collection of all the samples corresponding to the
$i$th frame. $\rv_i$ can be described by the following linear
model,
   \begin{eqnarray}
      \rv_i = \Sv_i\Av\bv_i + \nv_i
      \label{e. flat fadin received signal model}
   \end{eqnarray}
where $\Sv_i = [\sv_i^1\; \cdots \sv_i^K]$ is the matrix whose
columns are the spreading sequences used for spreading the $i$th
information symbol of all the users, that is $\sv_i^k
\defi [ s_{iN_fN_c+1}^k \;
s_{iN_fN_c+2}^k \; \cdots \; s_{(i+1)N_fN_c}^k]^T$; $\Av$ is a
diagonal matrix with the users' amplitudes on its diagonal, that
is $\Av = \frac{1}{\sqrt{N_f}}{\rm diag}[ \sqrt{E_1} \; \cdots \;
\sqrt{E_K}]$; $\bv_i = [b_i^1 \; \cdots b_i^K]^T$ is the vector
containing the $i$th transmitted symbols of all the users; and
$\nv_i \sim \N{0,\sigma^2\Iv}$ is the additive noise.  Note that a
coded IR systems can model an RCDMA system by taking $N_f = N$ and
letting $N_f$ be the processing gain of the RCDMA system. Also
note that as $N_f$ decreases, the transmitted signal become more
and more impulsive and hence the name Impulse Radio systems
\cite{Fishler:02a,Martret:00a,Sadler:02}.

It is well known that $\rv_i$ is a sufficient statistic for
detecting the transmitted symbols. It is also well known that
$\yv_i = \Sv_i^T\rv_i$ is also a sufficient statistic for
detecting the transmitted symbols. It is easily seen that $\yv_i$
can be described by the following model,
   \begin{eqnarray}
      \yv_i = \Rv_i\Av \bv_i + \tilde{\nv}_i
      \label{e. flat fading received signal model equivalent}
   \end{eqnarray}
where $\Rv_i = \Sv_i^T\Sv_i$ is the un-normalized
cross-correlation matrix, with $N_f$ on its main diagonal and
$\rho_{kl}^i = < \sv_i^{k}, \sv_i^l> = \sv_i^{kT}\sv_i^l =
\sum_{j=1}^{N}[\sv_i^k]_j[\sv_i^l]_j$, as the off-diagonal
elements; and where $\tilde{\nv}_i$ is a zero mean, Gaussian
random vector with correlation matrix $\sigma^2\Rv_i$.

Discrete time equivalent models for frequency selective channels
are more complex. In order to have a tractable model that allows
for analysis, we assume that a guard time equal to the length of
the channel impulse response exists at the end of each symbol.
This assumption is usually made in order to simplify the analysis
\cite{Evans:00}. As such, the following model for $\rv_i$ is used
in the sequel,
   \begin{eqnarray}
      \rv_i = \Hv_0\Sv_i\Av\bv_i + \nv_i
      \label{e. frequency selective fading received signal model}
   \end{eqnarray}
where $\Hv_0$ is a lower-triangular Toeplitz matrix whose first
column equals $[h_0\;h_1\; h_2 \; \cdots \; h_L\; \bfa{0}]^T$.

\subsection{Organization of the Paper}

The rest of the paper is organized as follows: In Section II, IR
systems transmitting over frequency-flat channels are analyzed,
and in Section III IR systems transmitting over frequency
selective channels are analyzed. In Section IV some conclusions
and concluding remarks are given. For convenience, numerical
examples are presented at the end of each section.

%%%%%%%%%%%%%%%%%%%%%%%%%%%%%%%%%%%%%%%%%%%%%%%%%%%%%%%%%%%%%%%%
%%%%%%%%%%%%%%%%%%%%%%%%%%%%%%%%%%%%%%%%%%%%%%%%%%%%%%%%%%%%%%%%
%%%%%%%%%%%%%%%%%%%%%%%%%%%%%%%%%%%%%%%%%%%%%%%%%%%%%%%%%%%%%%%%
% Two users, flat fading channel
%%%%%%%%%%%%%%%%%%%%%%%%%%%%%%%%%%%%%%%%%%%%%%%%%%%%%%%%%%%%%%%%
%%%%%%%%%%%%%%%%%%%%%%%%%%%%%%%%%%%%%%%%%%%%%%%%%%%%%%%%%%%%%%%%
%%%%%%%%%%%%%%%%%%%%%%%%%%%%%%%%%%%%%%%%%%%%%%%%%%%%%%%%%%%%%%%%

\section{Transmission Over Flat Fading Channels}
\label{s. two users flat}

In this section, we focus on systems transmitting over flat fading
channels. Both coded and uncoded systems are analyzed, and are
shown to behave differently as a function of the pulse rate.

%%%%%%%%%%%%%%%%%%%%%%%%%%%%%%%%%%%%%%%%%%%%%%%%%%%%%%%%%%%%%%%%
%%%%%%%%%%%%%%%%%%%%%%%%%%%%%%%%%%%%%%%%%%%%%%%%%%%%%%%%%%%%%%%%
% Coded System
%%%%%%%%%%%%%%%%%%%%%%%%%%%%%%%%%%%%%%%%%%%%%%%%%%%%%%%%%%%%%%%%
%%%%%%%%%%%%%%%%%%%%%%%%%%%%%%%%%%%%%%%%%%%%%%%%%%%%%%%%%%%%%%%%

\subsection{Coded System}

In this subsection coded-user systems are analyzed. The following
simple lemma will be very useful in the analysis of both the
matched filter detector and the optimal multiuser detector.
   \begin{Lemma}
      \label{l. asymptotic distribution of rho}
      Denote by $\rhov=[\rho_{1,2}\, \rho_{1,3}\, \ldots \,
      \rho_{K-1,K}]^T$, the vector containing cross-correlations between any
      two spreading sequences. Assume that $N \rightarrow \infty$ and that $\frac{N_f}{N_c}
      \rightarrow c>0$. Then
      $\rhov$ is asymptotically normally distributed with zero mean
      and correlation matrix $\frac{N_f}{N_c}\Iv$.
   \end{Lemma}
   \begin{PLemma}
      See Appendix \ref{a. asymptotic distribution rho}
   \end{PLemma}

Consider the model for the received signal (\ref{e. flat fading
received signal model equivalent}) scaled by
$\frac{1}{\sqrt{N_f}}$,
   \begin{eqnarray}
      \tilde{\yv}_i = N_f^{-1/2}\Rv_i \Av\bv + N_f^{-1/2}\nv.
   \end{eqnarray}

According to Lemma \ref{l. asymptotic distribution of rho},
asymptotically (for large $N$) the vector of elements of
$N_f^{-1/2}\Rv_i$ is distributed as a normal random vector with
zero mean and correlation matrix $\frac{1}{N}\Iv$. Thus, for
systems with large total processing gain, the distribution of
$\tilde{\yv}$, which is a sufficient statistic for detecting the
transmitted symbols, is essentially independent of the pulse rate,
and depends solely on $N$. Consequently the BER of any multiuser
detection (MUD) algorithm which is based on $\yv$ is essentially
independent of the pulse rate. In particular the performance of
the optimal, matched filter, minimum mean square error (MMSE), and
zero forcing (ZF) multiuser detectors are independent of the pulse
rate and depends solely on the total processing gain and not on
the ratio between the two types of processing gains. Simulation
results, which are not reported here due to space limitations,
show that this result holds for systems with processing gains as
low as $N=32$.

In the sequel we refer to a system satisfying $N_f << N$ as a low
pulse rate system, while a system such that $N_f$ is on the order
of $N$ is referred to as a high pulse rate system. RCDMA is an
example for a high pulse rate system since in this system $N_f =
N$. It should be noted that as the pulse rate, $N_f$, decreases,
the energy per transmitted pulse increases. This general behavior
characterizes the main hardware complexity tradeoff between high
pulse rate and low pulse rate systems. We demonstrate this
hardware complexity by examining the matched filter (MF) detector.

The MF detector for detecting the $i$th symbol of the first user
is $[\yv_i]_1 \lgth _{b_1 = -1}^{b_1 = 1} 0$. Denote by $R_s$ the
symbol rate. In order to implement the MF detector, the system
sampling rate can be as low as $N_fR_s$, while the transmitted
energy per pulse is $\frac{E}{N_f}$. These two terms represent a
hardware complexity tradeoff between high and low pulse rate
systems. While the sampling rate of low pulse rate systems can be
lower than the sampling rate used by high pulse rate systems, the
receiver dynamic range of low pulse rate systems must be higher
than the receiver dynamic range of high pulse rate systems. The
increase in the receiver dynamic range is due to the increase in
the signal peak-to-average power ratio exhibited by low pulse rate
systems.

%%%%%%%%%%%%%%%%%%%%%%%%%%%%%%%%%%%%%%%%%%%%%%%%%%%%%%%%%%%%%%%%

%%%%%%%%%%%%%%%%%%%%%%%%%%%%%%%%%%%%%%%%%%%%%%%%%%%%%%%%%%%%%%%%
%%%%%%%%%%%%%%%%%%%%%%%%%%%%%%%%%%%%%%%%%%%%%%%%%%%%%%%%%%%%%%%%
% Uncoded System
%%%%%%%%%%%%%%%%%%%%%%%%%%%%%%%%%%%%%%%%%%%%%%%%%%%%%%%%%%%%%%%%
%%%%%%%%%%%%%%%%%%%%%%%%%%%%%%%%%%%%%%%%%%%%%%%%%%%%%%%%%%%%%%%%

\subsection{Uncoded System}

In a way similar to the proof  of Lemma \ref{l. asymptotic
distribution of rho} it is easy to verify that in uncoded-user
systems $\rhov$ is asymptotically normally distributed with mean
$\frac{N_f}{N_c}$ and covariance matrix $\frac{N_f}{N_c}\(1 -
\frac{1}{N_c}\)\Iv$.

\subsubsection{Two User Systems}

{\em The Matched Filter Detector:} Consider a two-uncoded-user
system. For large N, the BER of the MF detector can be
approximated as follows:
   \begin{eqnarray}
      &&P_e = E_{\rho}(P_e|\rho) = E_{\rho}\left\{
      \frac{1}{2}Q\( \frac{ N_f \sqrt{\frac{E_1}{N_f}} -
      \sqrt{ \frac{E_2}{N_f} } \rho } {\sigma \sqrt{N_f} }\) \right. \nonumber\\
  &&+\left.
      \frac{1}{2}Q\( \frac{ N_f \sqrt{\frac{E_1}{N_f}} +
      \sqrt{ \frac{E_2}{N_f} } \rho } {\sigma \sqrt{N_f}  }\)
      \right\} \cong \nonumber\\
      &&\frac{1}{2}Q\( \frac{\sqrt{E_1} +
      \frac{\sqrt{E_2}}{N_c}}{\sqrt{\sigma^2 + \frac{E_2}{N}\(1 - \frac{1}{N_c} \)}} \)
      +
      \frac{1}{2}Q\( \frac{\sqrt{E_1} -
      \frac{\sqrt{E_2}}{N_c}}{\sqrt{\sigma^2 + \frac{E_2}{N}\(1 - \frac{1}{N_c} \)}} \)
      \label{e. Pe 2 users uncoded flat},
   \end{eqnarray}
where $P_e|\rho$ is due to \cite{Verdu:98}, and we used the
identity $\Eu{X}{Q\( \mu + \lambda X \)} = Q\(
\frac{\mu}{\sqrt{1+\lambda^2}} \)$ for $X\sim \N{0,1}$.

It can be easily seen that asymptotically (as
$N\rightarrow\infty$) the BER of the system depends on the ratio
between the two types of processing gain, and hence there is a
tradeoff between the two types of processing gains. The main
question that arises is ``what ratio between $N_f$ and $N_c$
minimizes the system BER?". In Appendix \ref{a. Analysis of the
Matched Filter receiver in uncoded system} the following result is
proven. Given that $\frac{E_1}{N}$ and $\frac{E_2}{N}$ are less
than $\sigma^2$, and that
$\frac{\sqrt{E_2}}{N}<\sqrt{E_1}<N\sqrt{E_2}$, the BER of (\ref{e.
Pe 2 users uncoded flat}) is a monotonically increasing function
of the pulse rate. The above sufficient conditions means that the
transmitted energy per chip is lower than the background noise
level, and that the energies transmitted by the two users do not
differ by a factor larger than the square of the processing gain.
These conditions are almost always met in practical systems.
Simulation results we conducted confirm that unless one of the
users is much stronger than the other, low pulse rate systems are
preferable over high pulse rate systems.

The superiority of low pulse rate systems over high pulse rate
system can be intuitively deduced from (\ref{e. Pe 2 users uncoded
flat}) quite easily. Let us assume that
$\frac{E_2}{N}<\frac{E_1}{N}<\sigma^2$. It can be easily seen that
under this condition the approximate BER of the MF detector is the
average of the $Q$ function over a simple random variable that can
take one of two possible values. Due to our assumptions, on one
hand, the average of these two values is approximately a constant
independent of the pulse rate, and on the other as the pulse rate
increases the distance between these two values increases as well.
Since the $Q$ function is a convex function, Jensen's inequality
implies that the BER of the MF detector is a monotonically
increasing function of the pulse rate.

%\subsubsection{Optimal Detector}
%\label{s. optimal flat two users uncoded}
{\em Optimal Detector:} In uncoded systems, it is quite clear from
the asymptotic distribution of the correlation between the two
users' spreading sequences that the distribution of any sufficient
statistic for detecting the transmitted symbols {\em depends} on
the pulse rate. Thus, it is of interest to study the BER of the
optimal MUD as a function of the pulse rate.

Denote by $P_e|\rho$ the probability of error of optimal MUD given
that the correlation between the two users' spreading sequences is
$\rho$. It is well known that no general closed form expression
for $P_e|\rho$ exists. Nevertheless, upper bounds for $P_e|\rho$
exist and it is easily seen that the one reported in
\cite{Verdu:98} is a monotonically increasing function of
$|\rho|$. As indicated by a large number of simulation studies, it
is widely believed that $P_e|\rho$ is a monotonically increasing
function of $|\rho|$ as well. Since in our system, $\rho$ is a
random variable, the overall BER is given by averaging $P_e|\rho$
with respect to the distribution of $\rho$. Denote by $P_e | N_f$
this overall BER of optimal MUD given that the pulse rate equals
$N_f$, i.e., $P_e| N_f = \Eu{\rho|N_f}{P_e|\rho}$.

Assume that $N_f < N_f'$. Note that since the system is an uncoded
one, $\rho | N_f \ge 0$ with probability one. In Appendix \ref{a.
monoticity of rho uncoded} it is proven that for all $x>0$ and
large $N$, $\Pr\( (\rho|N_f) < x \) \ge \Pr( (\rho|N_f') < x)$. In
the statistical literature this kind of relation between two
random variables is usually termed {\em first stochastic
dominance} and it is denoted as $\rho|N_f' \succeq_{FSD}
\rho|N_f$. It is well known that if $X \succeq_{FSD} Y$ and
$U(\cdot)$ is a monotonically increasing function then
$\Eu{X}{U(X)} \ge \Eu{Y}{U(Y)}$ \cite{Whitmore:82}. Thus since
$\rho|N_f' \succeq_{FSD} \rho|N_f$ and the upper bound for the
probability of error given $\rho$ is a monotonically increasing
function of $\rho$, the average upper bound, which is also an
upper bound for the average BER, is a monotonically increasing
function of $N_f$ as well. By using the conjecture that the
probability of error is a monotonically increasing function of
$\rho$, and by using $\rho|N_f' \succeq_{FSD} \rho|N_f$, we also
conjecture that the BER of the optimal MUD is a monotonically
increasing function of pulse rate.

\subsubsection{Multiple User Systems}

In this section the BER of uncoded-user systems with arbitrary
number of users using the matched filter detector or the optimal
detector is examined. It is demonstrated that the general behavior
observed for two-uncoded-user systems carries over to the case of
a large number of users.

The matched filter detector for detecting the transmitted symbol,
of say the first user, is, $[\yv_i]_1 = \sqrt{N_f E_1} b_1 +
\sum_{j=2}^K \sqrt{ \frac{E_j}{N_f}} b_j \rho_{1,j} + [\nv]_1
\lgth_{b_1 = -1}^{b_1 = 1} 0$, where $\rho_{1,j}$ is the $j$th
element of the first row of $\Rv_i$. Assuming that $K$ is large,
and all the users transmit with equal power, by invoking the
Central Limit Theorem (CLT) $\frac{1}{\sqrt{K-1}}\sum_{j=1}^K
\sqrt{ \frac{E_i}{N_f}} b_j \rho_{1,j}$ is asymptotically normally
distributed with zero mean and variance $\frac{E}{N_f}\(
\frac{N_f^2}{N_c^2} + \frac{N_f}{N_c}\( 1- \frac{1}{N_c} \) \)$.
As a result, the BER of the MF detector can be approximated by
   \begin{eqnarray}
      P_e \cong Q \( \frac{ \sqrt{E_1}}{\sqrt{ \sigma^2 + K E_2\(
      \frac{1}{N} + \frac{1}{N_c^2} - \frac{1}{N N_c} \) } } \).
   \end{eqnarray}
It is clear that the approximate system BER is a monotonically
increasing function of the pulse rate as is the case for two user
uncoded system. It is also clear that when the users transmit at
different powers similar conclusion can be reached.

Analyzing the performance of the optimal multiuser detector is
quite difficult due to the lack of closed-form expressions, or
even simple upper bounds, for the system BER as a function of the
users' gains and correlation matrix. Nevertheless, we conjecture
that, similarly to the case of a two-uncoded-user system, the
system BER is a monotonically increasing function of the pulse
rate. This conjecture is based on the observation that assuming
$N_f < N_f'$ then $\Rv_i | N_f \preceq_{FSD} \Rv_i | N_f'$. That
is, when the pulse rate decreases, the off-diagonal elements of
$\Rv_i$ tends to be smaller, and hence the spreading sequences of
the various users tends to be less correlated.

\subsection{Numerical Example}

\begin{figure}[ht]
   \includegraphics[scale=0.45]{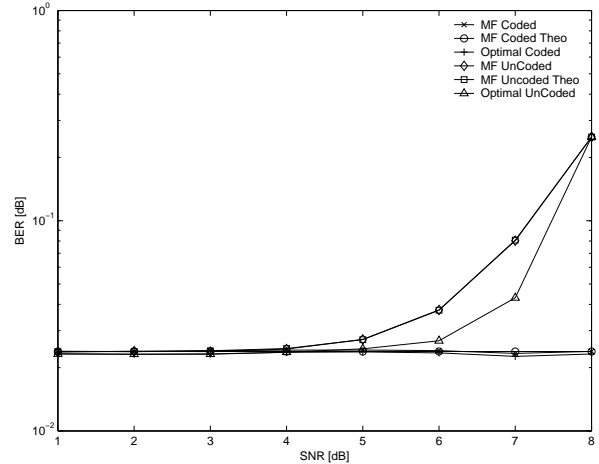}
   \caption{Probability of error as a function of the pulse rate.
   Two equal power users transmitting over a frequency-flat channel}
   \label{graph1a}
\end{figure}

\begin{figure}[ht]
   \includegraphics[scale=0.45]{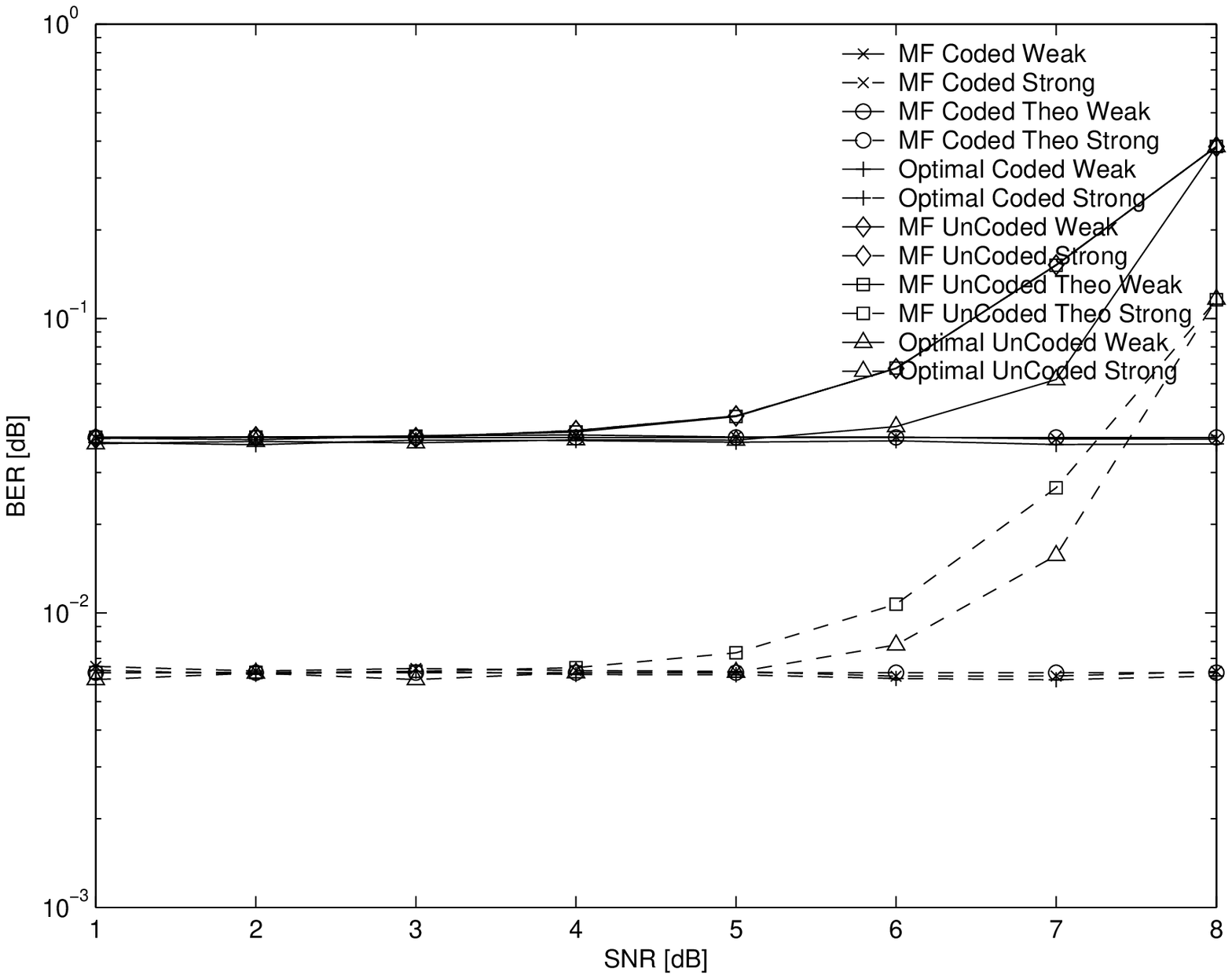}
   \caption{Probability of error as a function of the pulse rate.
   Two non-equal power users transmitting over a frequency-flat channel}
   \label{graph1b}
\end{figure}

In this subsection we present a numerical example that confirms
the results reported thus far. We consider a system with
processing gain $N=128$. Figures \ref{graph1a} and \ref{graph1b}
depict the BER of both the MF detector and the optimal MUD as a
function of the pulse rate. In Fig. \ref{graph1a} we assume two
equal power users transmitting at signal to noise ratio (SNR) of
6dB, while in Fig. \ref{graph1b} we assume that the first user
transmits at SNR of 5dB and the second use at SNR of 8dB. The
theoretical expressions for the performance of the matched filter
detector are depicted as well \cite{Fishler:02a}.

It is evident from the graph that the BERs of both the matched
filter detector and the optimal multiuser detector in the coded
system are unaffected by the pulse rate. Also the BERs of both the
matched filter detector and the optimal multiuser detector in the
uncoded system degrade considerably as the pulse rate increases.
This is in accordance with the analysis conducted in this section.
Moreover, we can see that the empirical and the theoretical curves
agree well.

%%%%%%%%%%%%%%%%%%%%%%%%%%%%%%%%%%%%%%%%%%%%%%%%%%%%%%%%%%%%%%%%%5
%%%%%%%%%%%%%%%%%%%%%%%%%%%%%%%%%%%%%%%%%%%%%%%%%%%%%%%%%%%%%%%%%5
%%%%%%%%%%%%%%%%%%%%%%%%%%%%%%%%%%%%%%%%%%%%%%%%%%%%%%%%%%%%%%%%%5
% Transmission over Frequency Selective channels
%%%%%%%%%%%%%%%%%%%%%%%%%%%%%%%%%%%%%%%%%%%%%%%%%%%%%%%%%%%%%%%%%5
%%%%%%%%%%%%%%%%%%%%%%%%%%%%%%%%%%%%%%%%%%%%%%%%%%%%%%%%%%%%%%%%%5
%%%%%%%%%%%%%%%%%%%%%%%%%%%%%%%%%%%%%%%%%%%%%%%%%%%%%%%%%%%%%%%%%5

\section{Transmission over Frequency Selective channels}
\label{s. two users isi channels}

%%%%%%%%%%%%%%%%%%%%%%%%%%%%%%%%%%%%%%%%%%%%%%%%%%%%%%%%%%%%%%%%%5
%%%%%%%%%%%%%%%%%%%%%%%%%%%%%%%%%%%%%%%%%%%%%%%%%%%%%%%%%%%%%%%%%5
% Analysis
%%%%%%%%%%%%%%%%%%%%%%%%%%%%%%%%%%%%%%%%%%%%%%%%%%%%%%%%%%%%%%%%%5
%%%%%%%%%%%%%%%%%%%%%%%%%%%%%%%%%%%%%%%%%%%%%%%%%%%%%%%%%%%%%%%%%5

\subsection{Analysis}
\label{ss. two users isi channels, analysis}

In this section, coded-user systems transmitting over frequency
selective channels are analyzed. The analysis will be carried out
in two stages. In the first stage it is assumed that only two
paths arrive at the receiver, that is, the channel impulse
response is $\hv=[1 \; \bfa{0}\; h_l]$, and that $l \le N_c$. In
the second step we will consider more general channels. Denote by
$r_j$ the sample at the output of the matched filter at the time
instant corresponding to the arrival time of the $j$th pulse from
the user of interest, say the first user. The following model for
$r_j$ can be easily deduced from (\ref{e. frequency selective
fading received signal model}),
   \begin{eqnarray}
      &&r_j = \sqrt{\frac{E_1}{N_f}} d_j^1 b_1 +
      \sqrt{\frac{E_1}{N_f}} h_l d_{j-1}^1 b_1 I_j^1 +
      \sqrt{\frac{E_2}{N_f}}d_j^2 b_2 I_j^2 \nonumber\\
      &&+
      \sqrt{\frac{E_2}{N_f}} h_l
      d_{j-1}^2 b_2 I_j^3 + n_j \quad, \quad j=1,\ldots, N_f,
      \label{e. received signal ISI two usrs}
   \end{eqnarray}
where $I_j^1$ is an indicator function taking the value one if the
$(j-1)$th pulse transmitted by the first user collides via the
second path with the $j$th pulse transmitted by the first user,
and zero otherwise. $I_j^2$ is a function taking the value one if
the $j$th pulse transmitted from the second user collides with the
$j$th pulse transmitted from the first user, the value $h_l$ if
the $j$th pulse transmitted from the second user and arriving via
the second path collides with the $j$th pulse transmitted from the
first user, and zero otherwise. $I_j^3$ is an indicator function
taking the value one if the $(j-1)$th pulse transmitted by the
second user collides via the second path with the $j$th pulse
transmitted by the first user, and zero otherwise.  The MF
detector is
   \begin{eqnarray}
      &&T = \sum_{j=1}^{N_f} d_j^1 r_j  = \sum_{j=1}^{N_f}
      \sqrt{\frac{E_1}{N_f}} b_1 +
      \sqrt{\frac{E_1}{N_f}} h_l d_j^1 d_{j-1}^1 b_1 I_j^1 \nonumber\\
      &&+
      \sqrt{\frac{E_2}{N_f}} d_j^1 d_j^2 b_2 I_j^2 \nonumber \\
      &&+
      \sqrt{\frac{E_2}{N_f}}
      h_l d_j^1d_{j-1}^2 b_2 I_j^3 + d_j^1 n_j
      \lgth_{b_1 = -1}^{b_1=1} 0.
   \end{eqnarray}
In Appendix \ref{a. asymptotic distribution of the interference
ISI} it is shown that the multiple access interference (MAI),
$\sum_{j=1}^{N_f} \sqrt{\frac{E_1}{N_f}} h_l d_j^1 d_{j-1}^1 b_1
I_j^1 + \sqrt{\frac{E_2}{N_f}} d_j^1 d_j^2 b_2 I_j^2 +
\sqrt{\frac{E_2}{N_f}} h_l d_j^1d_{j-1}^2 b_2 I_j^3$, is
asymptotically (as $N\rightarrow\infty$) normally distributed with
zero mean and variance $E_1\frac{h_l^2 l}{N_c^2} +
E_2\frac{1+h_l^2}{N_c}$. Thus for systems with large processing
gain, the distribution of the MF test statistic is approximately
$T \sim \N{ \sqrt{N_fE_1}b_1 , \frac{E_1 h_l^2 l}{N_c^2} +
E_2\frac{1+h_l^2}{N_c} + N_f\sigma^2 }$, and the BER can be
approximated by,
   \begin{eqnarray}
      P_e \cong Q\( \frac{ \sqrt{E_1} }{\sqrt{\sigma^2  + E_1\frac{h_l^2 l}{NN_c} +
            E_2\frac{1 + h_l^2}{N} } } \).
         \label{e. Pe 2 usrs ISI simple channel}
   \end{eqnarray}

It can be easily seen from Appendix \ref{a. asymptotic
distribution of the interference ISI} that the multiple access
interference is the sum of two independent terms. The first is the
MAI created by the second user, $\sqrt{\frac{E_2}{N_f}} d_j^1
d_j^2 b_2 I_j^2 + \sqrt{\frac{E_2}{N_f}} h_l d_j^1d_{j-1}^2 b_2
I_j^3$, and the second is the self interference processes created
by the first user upon itself, $\sum_{j=1}^{N_f}
\sqrt{\frac{E_1}{N_f}} h_l d_j^1 d_{j-1}^1 b_1I_j^1$. Note that
the self interference created by the first user is due to pulses
arriving via the second path colliding with different pulses
arriving via the first path; that is, this represents inter-pulse
interference.

We now turn to the computation of the BER of the MF when more than
two paths arrive at the receiver. The MAI created by the second
user is asymptotically (as $N\rightarrow\infty$) normally
distributed with zero mean and variance $E_2\frac{1+h_l^2}{N_c}$.
This MAI can be modeled as the sum of two independent, zero mean
Gaussian random variables, with variances $\frac{E_2}{N_c}$ and
$\frac{E_2 h_l^2}{N_c}$, respectively. The first (second) random
variable represents the MAI caused by pulses originating from the
second user and arriving at the receiver via the first (second)
path. Hence, asymptotically, the MAI resulting from pulses
arriving through different paths are independent. Using the same
method used in Appendix \ref{a. asymptotic distribution of the
interference ISI} this can be generalized to channels with more
than two paths or with two paths such that $l>N_c$. Thus, when the
channel impulse response is arbitrary, the MAI created by the
second user is asymptotically normally distributed with zero mean
and variance $\frac{E_2}{N_c} \sum_{i=1}^l h_i^2 =\frac{E_2
||\hv||}{N_c}$, which is the sum of the interference created by
the different paths.

If the number of users in the system is larger than two it is easy
to verify that the MAI processes due to different users are
independent. Hence the total multiple access interference due to
various users, is asymptotically (as $N\rightarrow\infty$)
normally distributed with zero mean and variance
$\frac{||\hv||}{N}\sum_{i=2}^K E_i$.

The self interference created by the first user due to pulses
arriving via the second path is asymptotically normally
distributed with zero mean and variance $\frac{E_2
h_l^2}{N_c}\frac{l}{N_c}$. It is easy to verify (Appendix \ref{a.
asymptotic distribution of the interference ISI}) that if $l>N_c$
the self interference caused by the first user is asymptotically
normally distributed with zero mean and variance $\frac{E_2
h_l^2}{N_c}$. It should be noted that when $l \le N_c$, the
average power of the self interference is the average power of the
self interference when $l>N_c$, $\frac{E_2 h_l^2}{N_c}$,
multiplied by the probability that a transmitted pulse will arrive
via the second path at times where the next transmitted pulse will
arrive via the first path. If more than two paths exist it can be
seen that the self interference terms created by any two paths are
asymptotically independent. Thus, similarly to the MAI created by
the second user, the total self interference created by the first
user is the sum of the individual interferences. The following
conclusion follows readily from this discussion: in arbitrary
channels, the self interference created by the first user is
asymptotically normally distributed with zero mean and variance
$\frac{E_1}{N_c}\( \sum_{j=1}^{N_c} \frac{j}{N_c}h_j^2 +
\sum_{N_c+1}^l h_j^2 \)$.

Combining the asymptotic distribution of the multiple access
interference and the self interference results in the following
simple approximate expression for the BER of the MF detector,
   \begin{eqnarray}
      P_e \cong Q\( \frac{ \sqrt{E_1} }{\sqrt{\sigma^2  + \frac{E_1}{N}\(
         \sum_{i=1}^{N_c} \frac{i}{N_c}h_i^2 + \sum_{N_c+1}^l
         h_i^2 \) + \frac{||\hv||}{N} \sum_{i=1}^K E_i }} \).
         \label{e. Pe ISI}
   \end{eqnarray}
Note that (\ref{e. Pe 2 usrs ISI simple channel}) is a special
case of (\ref{e. Pe ISI}).

It is very easy to see that the BER of the MF detector is
influenced by the pulse rate. The argument of the $Q(\cdot)$
function appearing in the expression for the BER, (\ref{e. Pe
ISI}), is a monotonically non-increasing function of $N_c$, or
equivalently, monotonically non-decreasing function the pulse
rate, $N_f$, and hence the BER is a monotonically non-decreasing
function of the pulse rate. Moreover, the effect of the pulse rate
on the BER is due to collisions between the transmitted pulses and
pulses received via the multipath from the user of interest. Thus,
as the pulse rate increases the probability of such collisions
increases as well, and hence the BER increases. On the other hand,
the interference caused by the other users is independent of the
pulse rate.

The result just obtained raises a question as to whether the pulse
rate has any effect on the performance of the (jointly or
individually) optimal multiuser detector. The answer to this
question is simple: a numerical example, discussed in the next
subsection, demonstrates that the pulse rate {\em does} effect the
BER of the system. This answer raises an even more interesting
question as to whether the pulse rate effects the BER in the same
way regardless of the scenario. The answer to this question is
much more complicated and the lack of a closed-form expression for
the BER of the optimal multiuser detector prohibits a definitive
answer. In the numerical example just mentioned, the BER of the
optimal MUD is a monotonically increasing function of the pulse
rate. We conjecture that performance improvement occurs with the
decrease of the pulse rate. Although we do not have mathematical
proof for this claim, we do have some evidence supporting it . By
invoking the CLT it could be argued that the mean of the
correlation matrix, $\Rv_i$, is independent of the pulse rate, and
that a decrease in the pulse rate uniformly ``concentrates" the
distribution of the correlation matrix around its mean; that is,
with some abuse of notation, $\sigma\( \Rv|N_f \) < \sigma( \Rv|
N_f')$ for $N_f < N_f'$. Combining this with the convexity of the
probability of error as a function of $\Rv$, provides evidence
supporting the conjecture. Numerous simulations we have conducted
supports this conjecture as well.

%%%%%%%%%%%%%%%%%%%%%%%%%%%%%%%%%%%%%%%%%%%%%%%%%%%%%%%%%%%%%%%%%5
%%%%%%%%%%%%%%%%%%%%%%%%%%%%%%%%%%%%%%%%%%%%%%%%%%%%%%%%%%%%%%%%%5
% Numerical Example
%%%%%%%%%%%%%%%%%%%%%%%%%%%%%%%%%%%%%%%%%%%%%%%%%%%%%%%%%%%%%%%%%5
%%%%%%%%%%%%%%%%%%%%%%%%%%%%%%%%%%%%%%%%%%%%%%%%%%%%%%%%%%%%%%%%%5

\subsection{Numerical Example}

In this subsection we present a numerical example that confirms
the results of section \ref{ss. two users isi channels, analysis}.
We consider a coded-system with processing gain $N=128$. For
simplicity, the channel was taken to be $\hv = [1 \;0.9\; 0.8]$.

\begin{figure}[ht]
   \includegraphics[scale=0.45]{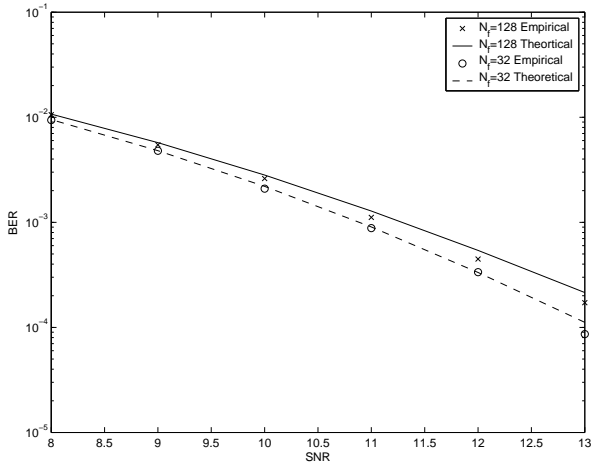}
   \caption{BER of the matched filter detector as a function of
   SNR, for an RCDMA system and a low pulse rate system}
   \label{graph2}
\end{figure}

Figure \ref{graph2} depicts the BER of the matched filter detector
as a function of the SNR. We consider a three-user system, one
user of which has SNR 3dB higher than the SNR of the other two.
The curves shown correspond to an RCDMA system, that is $N_f=
128$, and a system with pulse rate equal thirty two, that is $N_f
= 32$. The theoretical expressions for the BER are depicted as
well.

As can be seen from the figure, the BER of the lower pulse rate
system is lower than the BER of the RCDMA system, and a gain of
more than 0.5dB can be achived by using the low pulse rate system.
It is evident that the performance gap between the low pulse rate
and high pulse rate system increases as the SNR increases. Recall
that the self interference noise level increases as the pulse rate
increases (see, (\ref{e. Pe ISI}). Therefore as long as the
additive noise level and the MAI level are high compared with the
self interference level, then the difference in BER of high and
low pulse rate systems is negligible. However, when the SNR
increases and the self interference noise becomes dominant, then
the differences between low and high pulse rate systems become
evident.

\begin{figure}[ht]
   \includegraphics[scale=0.45]{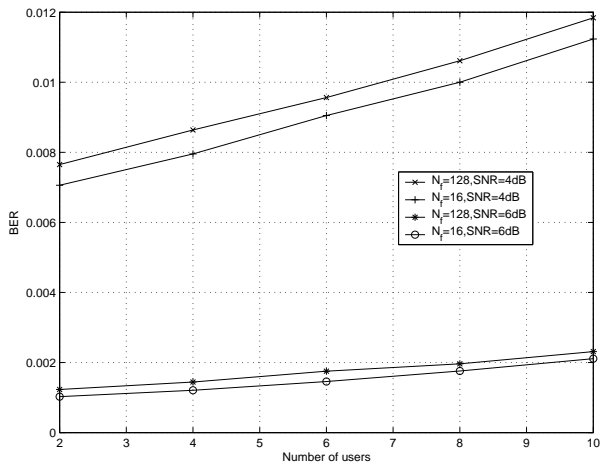}
   \caption{BER of the optimal MUD as a function of the number of users}
   \label{graph3}
\end{figure}

In Figure \ref{graph3} the optimal detector's BER is depicted as a
function of the number of users. We consider an RCDMA system and a
low pulse rate system transmitting 16 pulses per symbol, with
equal power users. We examined two SNRs 4dB and 6dB.

As can be seen from Fig \ref{graph3} the BER increases as the
number of users increases, and the low pulse rate system
outperforms the RCDMA system for any number of users. In addition,
it can be seen from the figure that for fixed SNR and BER the low
pulse rate system can support two additional users compared with
the RCDMA system.

In the last figure we examine the optimal detector's BER as a
function of the SNR and the pulse rate. We consider six equal
power users, and an RCDMA system and two low pulse rate systems
transmitting 32 and 8 pulses per symbol.

We can see from the figure that the RCDMA system requires an
additional 0.3dB to achieve the same BER as the low pulse rate
system. The performance improvement due to the use of low pulse
rate system is not large. However, this performance improvement
demonstrates the validity of our theoretical results as well asthe
advantages of using low pulse rate systems.

%In the next figure we depict the BER of the optimal detector as a
%function of the pulse rate. We consider a system with five equal
%power users with SNR of 5dB.

%%%%%%%%%%%%%%%%%%%%%%%%%%%%%%%%%%%%%%%%%%%%%%%%%%%%%%%%%%%%%%%%%
%%%%%%%%%%%%%%%%%%%%%%%%%%%%%%%%%%%%%%%%%%%%%%%%%%%%%%%%%%%%%%%%%
%%%%%%%%%%%%%%%%%%%%%%%%%%%%%%%%%%%%%%%%%%%%%%%%%%%%%%%%%%%%%%%%%
% Conclusions
%%%%%%%%%%%%%%%%%%%%%%%%%%%%%%%%%%%%%%%%%%%%%%%%%%%%%%%%%%%%%%%%%
%%%%%%%%%%%%%%%%%%%%%%%%%%%%%%%%%%%%%%%%%%%%%%%%%%%%%%%%%%%%%%%%%
%%%%%%%%%%%%%%%%%%%%%%%%%%%%%%%%%%%%%%%%%%%%%%%%%%%%%%%%%%%%%%%%%

\section{Summary and Concluding Remarks}

In this paper, the trade off between two types of processing gain,
namly spreading and time division, has been analyzed under the
assumption that the total processing gain is fixed and large.
These two types of processing gain are interchangeable, and the
analysis reveals that in some cases one should favor the second
type of processing gain over the first.

Specifically, it has been argued that when coded systems
transmitting over non-ISI channels are used, the two types of
processing gains are reciprocal. That is, the BER of many MUDs is
independent of the ratio between the two types of processing gain
as long as the total processing gain is fixed. Nevertheless, the
system complexity varies as the ratio between the two types of
processing gain is changed. In systems that devote some of their
processing gain to reducing the transmission time, the sampling
rate can be decreased at the expense of large dynamic range
requirements, when compared with high pulse rate systems.

In the context of UWB systems this result is very important. Under
today's regulations, the bandwidth of UWB systems could be up to
$7\;GHz$. It is obvious that RCDMA systems that use the whole
bandwidth will have to sample the received signal at rate of at
least $7\;GHz$. On the other hand, some of the processing gain can
be devoted to reducing the transmission time, and thus lower
sampling rate could be used. Moreover, multiuser detection
algorithms specifically designed for low pulse rate systems have
very low complexity  compared with their high pulse rate
counterparts \cite{Fishler:02}.

In frequency selective channels it has been shown that there is a
tradeoff between the two types of processing gain, and this
tradeoff is in favor of reducing the pulse rate, that is reducing
the total transmission time. Although the decrease in the BER due
to the use of low pulse rate systems can be low, the system
complexity will be much lower than that of high pulse rate
systems. It can be seen from the expression (\ref{e. Pe ISI}) for
the approximate BER of the MF detector that the effect of the
pulse rate on the total noise level is only via the signal
transmitted from the user of interest. If the number of users is
large or all the users transmit with equal power, the part of the
noise level depending on the pulse rate is negligible compared to
the part that is independent of the pulse level. In the
equal-power-users case it is easy to verify that the part
depending on the pulse rate is always smaller than the part
independent of the pulse rate. Therefore, in these cases the
advantage of the low pulse rate systems is negligible compared
with high pulse rate systems.

\bibliographystyle{plain}
\bibliography{uwb,mud,book}

\begin{appendices}

%\renewcommand{\baselinestretch}{1.85}
%\small\normalsize

%%%%%%%%%%%%%%%%%%%%%%%%%%%%%%%%%%%%%%%%%%%%%%%%%%%%%%%%%%%%
%%%%%%%%%%%%%%%%%%%%%%%%%%%%%%%%%%%%%%%%%%%%%%%%%%%%%%%%%%%%
%%%%%%%%%%%%%%%%%%%%%%%%%%%%%%%%%%%%%%%%%%%%%%%%%%%%%%%%%%%%
%%%%%%%%%%%%%%%%%%%%%%%%%%%%%%%%%%%%%%%%%%%%%%%%%%%%%%%%%%%%
% Proof of lemma 1
%%%%%%%%%%%%%%%%%%%%%%%%%%%%%%%%%%%%%%%%%%%%%%%%%%%%%%%%%%%%
%%%%%%%%%%%%%%%%%%%%%%%%%%%%%%%%%%%%%%%%%%%%%%%%%%%%%%%%%%%%
%%%%%%%%%%%%%%%%%%%%%%%%%%%%%%%%%%%%%%%%%%%%%%%%%%%%%%%%%%%%
%%%%%%%%%%%%%%%%%%%%%%%%%%%%%%%%%%%%%%%%%%%%%%%%%%%%%%%%%%%%

\section{Proof of Lemma \ref{l. asymptotic distribution of rho}}
\label{a. asymptotic distribution rho}

Recall that \\$\rhov = [\rho_{1,1}, \cdots , \rho_{K-1,K}]^T =
\sum_{j=1}^N [s_j^1s_j^2, \cdots , s_j^{K-1}s_j^K]^T =$\\
$\sum_{j=0}^{N_f-1}\sum_{l=1}^{N_c}[s_{jN_c+l}^1s_{jN_c+l}^2,
\cdots , s_{jN_c+l}^{K-1}s_{jN_c+l}^K]^T$. In order to prove the
lemma we first show that for $m\neq n$ the random vectors
$\sum_{l=1}^{N_c}[s_{mN_c+l}^1s_{mN_c+l}^2, \cdots ,
s_{mN_c+l}^{K-1}s_{mN_c+l}^K]^T$ and
$\sum_{l=1}^{N_c}[s_{nN_c+l}^1s_{nN_c+l}^2, \cdots ,
s_{nN_c+l}^{K-1}s_{nN_c+l}^K]^T$ are independent and identically
distributed with zero mean and covariance matrix equal
$\frac{1}{N_c}\Iv$. It is easy to see from the definition of
$s_j^k$ that, for every $k$, $s_j^k$ is independent of $s_l^k$ for
$l\neq \lfloor j/N_c \rfloor N_c+1, \ldots, \lceil j/N_c \rceil
N_c$. Thus, for $m\neq n$ the random variables $\{ s_{mN_c+l}^k
\}_{l=0,k=1}^{N_c-1,K}$ are jointly independent of $\{
s_{nN_c+l}^k \}_{l=0,k=1}^{N_c-1,K}$, and so the random vectors
$\sum_{l=1}^{N_c}[s_{mN_c+l}^1s_{mN_c+l}^2, \cdots ,
s_{mN_c+l}^{K-1}s_{mN_c+l}^K]^T$ and
$\sum_{l=1}^{N_c}[s_{nN_c+l}^1s_{nN_c+l}^2, \cdots ,
s_{nN_c+l}^{K-1}s_{nN_c+l}^K]^T$ are independent as well. We now
turn to examine the random variable
$\sum_{l=1}^{N_c}s_{jN_c+l}^ns_{jN_c+l}^m$, where $0<n<m\le K$ .
This sum will be equal to zero if the $j$th pulse of the $m$th and
$n$th users will be transmitted at different time slots. The
probability of this event is $1-1/N_c$. If the $j$th pulse of the
$m$th and $n$th users are transmitted at the same time slots, then
the probability that both of them will transmit a pulse with equal
phase is one-half. Combining these observations, it is easy to see
that $\sum_{l=1}^{N_c}s_{jN_c+l}^ns_{jN_c+l}^m$ is a ternary
random variable equaling zero with probability $1/N_c$, and one or
minus one each with probability $1-1/(2N_c)$. By using the same
technique one can verify that the expectation of
$\sum_{l=1}^{N_c}s_{jN_c+l}^ns_{jN_c+l}^m
\sum_{l=1}^{N_c}s_{jN_c+l}^ks_{jN_c+l}^l$ equals zero.

Now, $\rhov = \sum_{j=1}^{N_f} \Vv_j^T$, \\where $\Vv_j =
\sum_{l=1}^{N_c}[s_{jN_c+l}^1s_{jN_c+l}^2, \cdots ,
s_{jN_c+l}^{K-1}s_{jN_c+l}^K]^T$ is a zero mean random vector with
covariance matrix $\frac{1}{N_c}\Iv$. Invoking the CLT on this sum
proves the lemma; i.e.,
\begin{eqnarray}
   \sqrt{\frac{N_c}{N_f}}\rho \rightarrow N\(0,\Iv\)
\end{eqnarray}

%%%%%%%%%%%%%%%%%%%%%%%%%%%%%%%%%%%%%%%%%%%%%%%%%%%%%%%%%%%%
%%%%%%%%%%%%%%%%%%%%%%%%%%%%%%%%%%%%%%%%%%%%%%%%%%%%%%%%%%%%
%%%%%%%%%%%%%%%%%%%%%%%%%%%%%%%%%%%%%%%%%%%%%%%%%%%%%%%%%%%%
%%%%%%%%%%%%%%%%%%%%%%%%%%%%%%%%%%%%%%%%%%%%%%%%%%%%%%%%%%%%
%%%%%%% Performance of the match filter uncoded system
%%%%%%%%%%%%%%%%%%%%%%%%%%%%%%%%%%%%%%%%%%%%%%%%%%%%%%%%%%%%
%%%%%%%%%%%%%%%%%%%%%%%%%%%%%%%%%%%%%%%%%%%%%%%%%%%%%%%%%%%%
%%%%%%%%%%%%%%%%%%%%%%%%%%%%%%%%%%%%%%%%%%%%%%%%%%%%%%%%%%%%
%%%%%%%%%%%%%%%%%%%%%%%%%%%%%%%%%%%%%%%%%%%%%%%%%%%%%%%%%%%%

\section{Analysis of the Matched Filter receiver in uncoded system}
\label{a. Analysis of the Matched Filter receiver in uncoded
system}

Define two functions $f_1(N_c),f_2(N_c)$ as follow,
\begin{eqnarray}
   f_1(N_c) = \frac{\sqrt{E_1} +
   \frac{\sqrt{E_2}}{N_c}}{\sqrt{\sigma^2 + \frac{E_2}{N}\( 1 -
   \frac{1}{N_c}\)}} \quad ; \quad f_2(N_c) = \frac{\sqrt{E_1} -
   \frac{\sqrt{E_2}}{N_c}}{\sqrt{\sigma^2 + \frac{E_2}{N}\( 1 -
   \frac{1}{N_c}\)}}.
   \label{def. auxfun}
\end{eqnarray}

The bit error rate (\ref{e. Pe 2 users uncoded flat}) can be
easily expressed with the aid $f_1(N_c),f_2(N_c)$, and it is given
by $P_e = 1/2Q\( f_1(N_c) \) + 1/2Q\( f_2(N_c) \)$. In what
follows we find sufficient conditions such that the BER is a
monotonically decreasing function of $N_c$, which is equivalent to
proving that the BER is a monotonically increasing function of the
pulse rate. In order to prove that the BER is a monotonically
decreasing function of $N_c$ we first take the derivative of the
BER with respect to $N_c$, which after some manipulation can be
seen to be given by
\begin{eqnarray*}
   &&\frac{\partial P_e}{\partial N_c} = \frac{\partial Q(f_1(N_c))
   + Q(f_2(N_c))}{\partial N_c}
   \nonumber\\
   &&=
   e^{-f_1^2(N_c)}\( \sigma^2 + E_2N^{-1} - 1/2E_2 N^{-1}N_c^{-1} +
   1/2\sqrt{E_1E_2}N^{-1}
   \)\\
   &&-e^{-f_2^2(N_c)}\( \sigma^2 + E_2N^{-1} - 3/2E_2 N^{-1}N_c^{-1} +
   1/2\sqrt{E_1E_2}N^{-1}
   \).
\end{eqnarray*}

In order for the BER to be a monotonically decreasing function of
$N_c$, the derivative of the BER should be negative for every
$N_c$. This is equivalent to the following condition,
   \begin{eqnarray}
      &&f_2^2(N_c) - f_1^2(N_c)\nonumber\\
      &&< \ln \( \frac{ \sigma^2 + E_2N^{-1} - 3/2E_2 N^{-1}N_c^{-1} 
   1/2\sqrt{E_1E_2}N^{-1}}{\sigma^2 + E_2N^{-1} - 1/2E_2 N^{-1}N_c^{-1} +
   1/2\sqrt{E_1E_2}N^{-1}} \) \nonumber\\
 &&\quad,\quad \forall N_c.
   \label{t101}
   \end{eqnarray}
Assume that $\frac{E_1}{N},\frac{E_2}{N}<\sigma^2$. Substituting
(\ref{def. auxfun}) into (\ref{t101}), and by using our
assumption, the following sufficient condition is deduced,
   \begin{eqnarray}
      &&\frac{4\sqrt{E_1E_2}}{N_c\sigma^2+\frac{N_cE_2}{N}\(1-\frac{1}{N_c}\)} 
      \nonumber\\
      &&>
      \ln \( \frac{ \sigma^2 + E_2N^{-1} - 1/2E_2 N^{-1}N_c^{-1} }
      {\sigma^2 + E_2N^{-1} - 3/2E_2 N^{-1}N_c^{-1} }\) \quad, \forall N_c.
      \label{t102}
   \end{eqnarray}
Upper bounding the right-hand side using the bound $\ln(1+x)<x$
results in the following sufficient condition,
\begin{eqnarray}
      \frac{4\sqrt{E_1}}{\sigma^2+\frac{E_2}{N}\(1-\frac{1}{N_c}\)} >
       \frac{ \sqrt{E_2} N^{-1} }
      {\sigma^2 + E_2N^{-1} - 3/2E_2 N^{-1}N_c^{-1} } , \forall N_c.
      \label{t103}
   \end{eqnarray}
Bounding the denominator of the left-hand side from above by
$2\sigma^2$, and the denominator of the right hand side from below
by $\sigma^2/2$, results in the following sufficient condition:
$\sqrt{E_1}
> \frac{\sqrt{E_2}}{N}$.

%%%%%%%%%%%%%%%%%%%%%%%%%%%%%%%%%%%%%%%%%%%%%%%%%%%%%%%%%%%%
%%%%%%%%%%%%%%%%%%%%%%%%%%%%%%%%%%%%%%%%%%%%%%%%%%%%%%%%%%%%
%%%%%%%%%%%%%%%%%%%%%%%%%%%%%%%%%%%%%%%%%%%%%%%%%%%%%%%%%%%%
%%%%%%%%%%%%%%%%%%%%%%%%%%%%%%%%%%%%%%%%%%%%%%%%%%%%%%%%%%%%
%%%%%%% Performance of the match filter uncoded system
%%%%%%%%%%%%%%%%%%%%%%%%%%%%%%%%%%%%%%%%%%%%%%%%%%%%%%%%%%%%
%%%%%%%%%%%%%%%%%%%%%%%%%%%%%%%%%%%%%%%%%%%%%%%%%%%%%%%%%%%%
%%%%%%%%%%%%%%%%%%%%%%%%%%%%%%%%%%%%%%%%%%%%%%%%%%%%%%%%%%%%
%%%%%%%%%%%%%%%%%%%%%%%%%%%%%%%%%%%%%%%%%%%%%%%%%%%%%%%%%%%%

\section{Properties of the Correlation Coefficient in an Uncoded
Systems} \label{a. monoticity of rho uncoded}

In this appendix the correlation between the spreading sequences
of two uncoded users is examined. In particular, it is proven
that, asymptotically in $N$, $\rho|N_f \preceq_{FSD} \rho| N_f'$
for $1 \le N_f < N_f' \le N$, or equivalently that $\rho|N_c
\succeq_{FSD} \rho| N_c'$ for $1 \le N_c < N_c' \le N$, where
$\preceq_{FSD}$ denotes first stochastic domination.

In order to prove that $\rho|N_c' \preceq_{FSD} \rho| N_c$ we have
to prove that for a fixed $x$, the probability of the event $\{
(\rho|N_c) \le x \}$ is a monotonically increasing function of
$N_c$. Recall that the correlation between the spreading sequences
given $N_c$, $\rho|N_c$, is asymptotically normally distributed
with mean $\frac{N_f}{N_c}$, and variance
$\frac{N_f}{N_c}\(1-\frac{1}{N_c}\)$. Thus, the probability of the
event $\{ (\rho|N_c) < x \}$ is, asymptotically, $\Pr\( (\rho|N_f)
< x \) = 1 - Q\( \frac{ x -
      N_f/N_c }{\sqrt{N_f/N_c\(1-
      1/N_c\)}} \)$. After some manipulation and using the relation
$N_fN_c=N$, the asymptotic probability of error is thus given by
   \begin{eqnarray}
      \Pr\( (\rho|N_c) < x \) = 1 - Q\( \frac{ x N_c^2 -
      N}{\sqrt{N N_c \(N_c - 1 \)}} \).
      \label{a. e. probability2}
   \end{eqnarray}

Assume that $N_c>1$. In order to prove that (\ref{a. e.
probability2}) is a monotonically increasing function of $N_c$, it
suffices to prove that the argument of the $Q$ function in
(\ref{a. e. probability2}) is a monotonically increasing function
of $N_c$. Differentiating the argument of the $Q$ function in
(\ref{a. e. probability2}) with respect to $N_c$, and omitting
some positive scaling factors, results in the following,
   \begin{eqnarray}
      &&\frac{\partial}{\partial N_c} \frac{ x N_c^2 -
      N}{\sqrt{N N_c \(N_c - 1 \)}} \nonumber\\
    &&= 4x N N_c^2(N_c-1) -
      \(x N_c^2-N\)\(N(N_c-1) + N N_c\) \nonumber\\
      &&=x N_c^2\( 2N_c
      - 3 \) + N(2N_c - 1).
      \label{a. e. probability3}
   \end{eqnarray}
It is easy to see that for $x \ge 0$, (\ref{a. e. probability3})
is positive, and hence $\frac{ x N_c^2 - N}{\sqrt{N N_c \(N_c - 1
\)}}$ is a monotonically increasing function of $N_c$. Thus $\Pr\(
(\rho|N_c) < x \)$ is a monotonically increasing function of $N_c$
as well. Note that since the system is an uncoded system, $\Pr\( (
\rho | N_c < 0) \) =0$, so the case $x < 0$, has no interest.

Assume that $N_c=1$. It is easy to verify that $\Pr\( ( \rho |
N_c=1 ) < x \) =0$ for $x < N_f$ and $\Pr\( ( \rho | N_c=1 ) < N_f
\)=1$. Combining this result with the monotonicity of $\Pr\( (
\rho | N_c ) < \alpha \)$ for $N_c > 1$, concludes the proof.

\section{Properties of the interference}
\label{a. asymptotic distribution of the interference ISI}

It is easy to verify from the definition of $d_i^j$ that the
random variables $d_j^1d_{j-1}^1,d_j^1d_j^2,\mbox{ and }
d_j^1d_{j-1}^2$ are binary random variables taking each of the
values $\pm 1$ with probability 1/2. Let us derive the
distribution of the random variable $h_l d_i^1d_{i-1}^1I_1$. The
$(j-1)$th pulse transmitted by the first user will arrive via the
second path at times when the $j$th pulse transmitted by the first
user might be received if and only if $\prod_{m=1}^{N_c-l}
s_{(j-1)N_c+m}^1 = 0$, and the probability of this event is
$\frac{l}{N_c}$. Given that $\prod_{m=1}^{N_c-l} s_{(j-1)N_c+m}^1
= 0$ the probability that a collision will occur is
$\frac{1}{N_c}$ (the probability of that arrival time of the $j$th
pulse via the main path is equal to the arrival time of the
$(j-1)$th pulse via the second path). Thus the probability of the
event $\{ I_j^1 = 1\}$ is $\frac{l}{N_c^2}$. Combining the
distribution of $d_j^1d_{j-1}^1$ and the distribution of $I_j^1$
results in the following distribution for $h_l d_j^1 d_{j-1}^1
I_j^1$: $h_l d_j^1 d_{j-1}^1 I_j^1$ can take on the values
$h_l,0,-h_l$ with probabilities
$1/(2N_c^2),1-1/(N_c^2),1/(2N_c^2)$ respectively. It is easy to
show in a similar way that the marginal distribution of $h_l
d_j^1d_{j-1}^2I_j^3$ is equal to the marginal distribution of $h_l
d_j^1 d_{j-1}^1 I_j^1$.

Similar arguments can lead to the following marginal distribution
of $d_j^1d_j^2I_j^2$: $d_j^1d_j^2I_j^2$ takes on the values
$1,h_l,0,-h_l,-1$ with probabilities $1/(2N_c), 1/(2N_c) -
l/(2N_c^2), 1 - 2/N_c + j/N_c^2, 1/(2N_c) - l/(2N_c^2), 1/(2N_c)$,
respectively. It is easy to verify that $\{ h_l
d_j^1d_{j-1}^1I_j^1 \}$, $\{d_j^1d_j^2I_j^3\}$, $\{h_l
d_j^1d_{j-1}^2I_j^3\}$ are zero mean white, mutually uncorrelated
random sequences. For example take $h_l d_j^1d_{j-1}^1I_j^1$, and
let $j$ and $k$ be two distinct time indices. The mean of $h_l
d_j^1d_{j-1}^1I_j^1 \cdot h_l d_k^1d_{k-1}^1I_k^1$ is
   \begin{eqnarray}
      \E{ h_l^2 d_j^1d_{j-1}^1I_j^1 d_k^1d_{k-1}^1I_k^1}
      =\E{d_j^1}\E{ h_l^2 d_{j-1}^1I_j^1 d_k^1d_{k-1}^1I_k^1} = 0
   \end{eqnarray}
where for the last equality we used the fact the $d_j^1$ is
independent of all the other random variables, and we assumed that
$j\neq k-1$ (if $j=k-1$, take $d_{k-1}^1$ instead).

The total interference is given by \\$\sum_{j=1}^{N_f}\left[
\sqrt{\frac{E_1}{N_f}}h_l d_j^1d_{j-1}^1I_j^1 +
\sqrt{\frac{E_2}{N_f}} d_j^1d_j^2I_j^2 + \sqrt{\frac{E_2}{N_f}}
h_l d_j^1d_{j-1}^2I_j^3\right]$. Since the three random processes
$\{ h_l d_j^1d_{j-1}^1I_j^1\}$, $\{d_j^1d_j^2I_j^3\}$, $\{ h_l
d_j^1d_{j-1}^2I_j^3 \}$ are zero mean and mutually uncorrelated,
the mean and the variance of $\sqrt{\frac{E_1}{N_f}}h_l
d_j^1d_{j-1}^1I_j^1 + \sqrt{\frac{E_2}{N_f}} d_j^1d_j^2I_j^2 +
\sqrt{\frac{E_2}{N_f}} h_l d_j^1d_{j-1}^2I_j^3$ are zero and
$\frac{E_1}{N_f} \frac{ h_l^2 l}{N_c^2}  + \frac{E_2}{N_f}
\frac{1+h_l^2}{N_c}$, respectively. Although $\left\{
\sqrt{\frac{E_1}{N_f}} h_ld_j^1d_{j-1}^1I_j^1 +
\sqrt{\frac{E_2}{N_f}} d_j^1d_j^2I_j^2 + \sqrt{\frac{E_2}{N_f}}
h_l d_j^1d_{j-1}^2I_j^3 \right\}$ is a white random sequence, it
is not an independent one. Nevertheless, it is a $1$-dependent
random sequence, and hence it is a $\phi$-mixing random sequence
for which the conditions in \cite{Bilingsley:86} hold. Thus, a
central limit theorem can be invoked, implying the asymptotic
normality of the total interference,
   \begin{eqnarray}
      &&\sum_{j=1}^{N_f} \sqrt{\frac{E_1}{N_f}} h_l
       d_j^1d_{j-1}^1I_i^1 +
      \sqrt{\frac{E_2}{N_f}} d_j^1d_j^2I_j^2 \nonumber\\
      &&+
      \sqrt{\frac{E_2}{N_f}} h_l d_j^1d_{j-1}^2I_j^3
      \sim \N{0, \frac{2 E_1 h_l^2 l}{N_c^2} +
      E_2\frac{1+h_l^2}{N_c}}.
   \end{eqnarray}

\end{appendices}

\begin{figure}[ht]
   \includegraphics[scale=0.45]{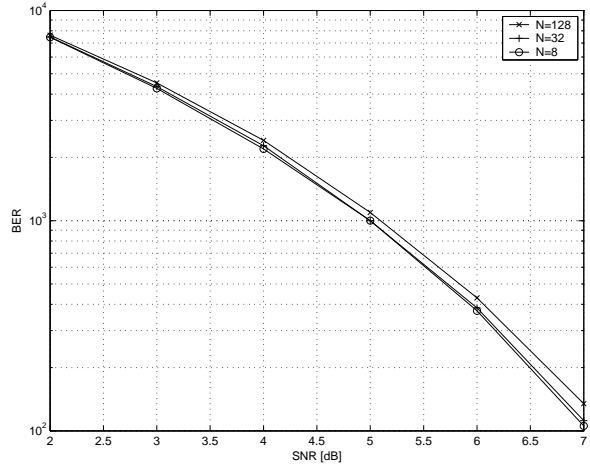}
   \caption{BER of the optimal MUD as a function of the SNR}
   \label{graph4}
\end{figure}

\end{document}